\title{Bayesian Variable Selection for Cox Regression Model with Spatially Varying Coefficients with Applications to Louisiana Respiratory Cancer Data}
\author{Jinjian Mu, Qingyang Liu, Lynn Kuo, Guanyu Hu}
\date{}

\documentclass[12pt]{article}
\usepackage[margin=1in]{geometry}
\usepackage{graphicx}
\usepackage{amsmath,amsthm,amssymb,bm}
\usepackage{float}
\usepackage{mathrsfs}
\usepackage{amsmath}
\usepackage{algorithm}
\usepackage{algorithmic}
\usepackage[authoryear]{natbib}
\usepackage{listings}
\usepackage{xcolor}
\usepackage{bm}
\usepackage{xcolor}
\usepackage{multirow}
\usepackage{xcolor}
\usepackage{hyperref}

\newcommand{\ntt}{\normalfont\ttfamily}

\newcommand{\pkg}[1]{{\protect\ntt#1}}

\begin{document}

\maketitle

\begin{abstract}
The Cox regression model is a commonly used model in survival analysis. In public health studies, clinical data are often collected from medical service providers of different locations. There are large geographical variations in the covariate effects on survival rates from particular diseases. In this paper, we focus on the variable selection issue for the Cox regression model with spatially varying coefficients. We propose a Bayesian hierarchical model which incorporates a horseshoe prior for sparsity and a point mass mixture prior to determine whether a regression coefficient is spatially varying. An efficient two-stage computational method is used for posterior inference and variable selection. It essentially applies the existing method for maximizing the partial likelihood for the Cox model by site independently first, and then applying an MCMC algorithm for variable selection based on results of the first stage. Extensive simulation studies are carried out to examine the empirical performance of the proposed method. Finally, we apply the proposed methodology to analyzing a real data set on respiratory cancer in Louisiana from the SEER program.

\end{abstract}

\section{Introduction}
In survival analysis, most studies focus on the overall effects of the covariates regardless of the spatial variation. In these models, associations between covariates and health outcomes are assumed to be constant. However, the effects of some covariates, such as pollution, education, employment status, may vary across different locations. Thus, models allowing for spatially varying covariate effects can be more flexible and more powerful to detect the relationship between covariates and health outcomes. From Tobler's first law of geography \citep{tobler1970computer}, ``Everything is related to everything else, but near things are more related than distant things,'' we know that the effects of covariates may be more similar to those in nearby regions than to those from distant regions due to similar environmental circumstances. Most existing literature \citep{banerjee2005semiparametric,zhou2008joint,zhou2015modelling,zhang2016bayesian} focus on adding spatial random effects as a separate spatial component to survival models. \cite{gelfand2003spatial} proposed a spatially varying coefficient model 
which incorporates the spatial random effects by assuming the regression coefficients follow a spatial process. \cite{reich2010bayesian} extended Gelfand's work in the generalized linear model framework with a spike-and-slab prior. Recently, \cite{hu2020modified} proposed modified versions of the Kaplan–Meier and Nelson–Aalen estimators which can represent the local survival curve and cumulative hazard. \cite{xue2019geographically} proposed a geographically weighted Cox regression model to analyze geographically distributed survival data. But they mainly added the geographical weights to each observation without assuming any probability model on spatially varying coefficients. \cite{hu2017spatial} proposed a parametric accelerated failure time (AFT) model with spatially varying coefficients. \cite{geng2020bayesian} explored spatial heterogeneity patterns of baseline hazards and regression coefficients. However their work was mainly concerned about the coefficient estimation and density estimation for survival models. Motivated by these works, we propose a Bayesian approach for variable selection for the Cox regression model \citep{cox1972regression} with spatially varying coefficients. Essentially, we add spatial correlation structure and variable selection formulation to the above mentioned spatially varying coefficient model \citep{gelfand2003spatial}.

In the Bayesian framework, many studies already proposed Bayesian variable selection methods for Cox models without spatially varying coefficients. \cite{ibrahim1999bayesian} developed a semi-automatic Bayesian variable selection method for up to 20 covariates and \cite{lee2011bayesian} built a penalized semiparametric method for high dimensional survival data. In spatial statistics, some Bayesian variable selection approaches are also well-developed. \cite{reich2010bayesian} proposed an approach for variable selection in multivariate spatially varying coefficient regression and applied a stochastic search algorithm to determine the probabilities that each covariate's effect is null, non-null but stationary across space, or spatially varying. \cite{choi2018bayesian} developed a flexible spatial variable selection method based on the Kuo-Mallick Entry Parameter (KMEP) method by \cite{kuo1998variable}. In recent years, horseshoe prior \citep{carvalho2009handling, carvalho2010horseshoe} gained much attention as a shrinkage-based variable selection method because it was shown to be robust to unknown sparsity patterns and to large outlying signals. The authors also suggest it as a good default prior. We believe it can be used widely because of its robust performance. We also incorporate this prior in our proposed model.

Our model is essentially a hierarchical Bayesian model with 3 levels: The first level is a Cox proportional hazards model \citep{cox1972regression} with site specific spatially varying coefficients. The second level is built on each component of the regression coefficients across all sites. Here, we extend the horseshoe prior in \cite{carvalho2009handling, carvalho2010horseshoe} by adding a spatial correlation matrix for variable selection. The third level is the prior belief on the hyperparameters in the second level. On estimation, we first employ the usual techniques for Cox models to maximize the partial likelihood independently for each site using \textsf{R} \pkg{survival} package. Then we model the obtained regression coefficients for each component using a multivariate normal distribution. Following the same settings in levels 2 and 3 of the original hierarchical model, we develop an MCMC algorithm for variable selection. We conduct simulation studies to illustrate our method and further apply our method to the Louisiana respiratory cancer data, which is downloaded from the Surveillance, Epidemiology, and End Results (SEER) Program.

The major contribution of this paper is to combine the horseshoe prior and spatial correlation matrix to propose a ``two-stage'' variable selection method for the Cox regression model with spatially varying coefficients.
Horseshoe prior, which is known to be effective in handling sparsity, is a shrinkage-based variable selection prior where the selection can be achieved by setting a threshold for the local shrinkage parameter.
Moreover, we add a spatial correlation matrix to the horseshoe prior and use a point mass mixture prior to further distinguish the spatially varying coefficients from spatially static coefficients. In the famous BYM model \citep{besag1991bayesian}, the authors defined the precision matrix for the spatial random effect by the adjacency matrix and a diagonal matrix. The horseshoe prior can be applied to the precision matrix. However, we adopt the idea in \cite{reich2010bayesian} and use the distance matrix to define the spatial correlation (covariance) matrix, which can be spatially more informative. The second contribution of this paper is on the estimation method. Instead of running an MCMC sampler based on the original model, we first estimate the Cox model for each site independently and then use the estimates from the partial likelihood to carry out an MCMC algorithm.
This method is more efficient in computation than the one based on the original model.

The rest of the paper is organized as follows. In Section~\ref{sec:prelim}, we discuss preliminary work for spatial varying models and variable selection models. In Section~\ref{sec:cox}, we introduce the proposed model with its hierarchical structure. Section~\ref{sec:est} shows the two-stage computing method and the choice of hyperprior. We conduct three simulation studies with various degree of sparsity and spatial signal in Section~\ref{sec:simulation} and illustrate our proposed method by a real data analysis on a SEER cancer study in Section~\ref{sec:realdata}. Some discussions about our method are presented in Section~\ref{sec:discussion}.

\section{Preliminary Work}\label{sec:prelim}

We first discuss some preliminary work on spatially varying coefficient models, Bayesian variable selection and graph distance that are needed for our approach.

\subsection{Spatially Varying Coefficients Model}

Let $Y(s)$ be a response function evaluated at the site $s$. We can write the usual Gaussian stationary spatial process model as in, for example, \cite{cressie1992statistics},
\begin{equation}
	Y(s)=\mu(s)+W(s)+\epsilon(s)
	\label{spatialmod}
\end{equation}
where $\mu(s)=\bm{x}(s)^T\bm{\beta}$ is the location mean effect, $\epsilon(s)\sim N(0,\tau^2)$ is a white noise process, and $W(s)$ is a second-order stationary mean 0 process independent of the white noise process. To be specific, we assume $E(W(s))=0$, $var(W(s))=\sigma^2$, and $cov(W(s),W(s'))=\sigma^2\rho(s,s';\phi)$, where $\rho$ is a valid two-dimensional correlation function. Here $W(s)$ can be treated as spatial random effects.

For linear models, \eqref{spatialmod} can be written as a spatially varying coefficient model \citep{gelfand2003spatial} as following:
\begin{eqnarray}
	Y(s)=\bm{X}^T(s)\bm{\tilde{\beta}}(s)+\epsilon(s)
	\label{SVC}
\end{eqnarray}
where $\bm{X}(s)$ is the covariate matrix at site $s$ and $\bm{\tilde{\beta}}(s)$ is assumed to follow a $p$-variate spatial process model.

\subsection{Bayesian Variable Selection with Shrinkage Prior}
We first discuss a general Bayesian variable selection method ignoring the site effect. Let $\bm{\beta}$ denote a $p$-dimensional coefficients ($\beta_1,...,\beta_p)^T$. An easy way to incorporate sparsity in a Bayesian regression problem is to use a point mass mixture prior for each component:
\begin{equation}
	\beta_k \sim (1-\pi)\delta_0+\pi g_{\beta_k},\qquad k=1,2,...,p,
	\label{pointmass prior}
\end{equation}
where $\pi=\text{Pr}(\beta_k\neq 0)$, and $g_{\beta_k}$ is a continuous density. And we can put a beta prior on $\pi$ to construct a Beta-Bernoulli prior on the model. There are a lot of computational issues due to the discontinuity of point mass mixture priors. Instead, many authors introduced continuous shrinkage priors which can be generally represented as global-local mixtures of Gaussian distributions:
\begin{equation}
	\beta_k \sim N(0,\tau^2\lambda_k^2),\qquad \tau\sim g, \qquad \lambda_k \sim f,
	\label{GL mixturess}
\end{equation}
where $\tau$ controls the global shrinkage and $\lambda_k$ controls the local shrinkage only on $\beta_k$. The global-local shrinkage priors have computational advantages over other variable selection priors, because the normal priors allow for conjugate updating of $\beta_k$ and $\lambda_k$, although the global-local shrinkage priors do not allow $\beta_k$'s to be exactly zero. There are many types of Bayesian shrinkage priors including Bayesian lasso \citep{park2008bayesian}, relevance vector machine \citep{tipping2003relevance}, normal-gamma mixtures \citep{griffin2010inference} and the horseshoe prior  \citep{carvalho2009handling, carvalho2010horseshoe}. In horseshoe prior, $f$ in \eqref{GL mixturess} is set as a standard half-Cauchy distribution $\text{Ca}^+(0,1)$, and usually $g$ is also a standard half-Cauchy distribution. For unknown sparsity patterns and large outlying signals, the horseshoe prior gives more robust results. 
We will apply \eqref{GL mixturess} in the form of the horseshoe prior for variable selection in this paper. 

\subsection{Graph Distance}
In this paper, we choose the graph distance $d(\cdot,\cdot)$ as the distance measure, similar to that in \cite{xue2019geographically}. We can define a graph $G$ by its nodes $V(G)= \{v_1, \ldots, v_n\}$ and connected with edges $E(G)=\{e_1, \ldots, e_m\}$. Then the graph distance between two nodes $v_i$ and $v_j$ can be defined as:
\begin{equation}
  d(v_i, v_j) =
    \begin{cases}
      \min |P| & \text{if $v_i$ and $v_j$ are connected by a path $P$ constructed from the edges $E(G)$,}\\
      \infty & \text{if $v_i$ and $v_j$ are not connected,}\\
    \end{cases}
    \label{graphdist}
\end{equation}
where $|P|$ is the number of edges in the path. We can regard the spatial structure of Louisiana state as a graph with every county as a vertex, and there is an edge between two counties if they are adjacent. In this way, the graph distance can be calculated for every two counties using \eqref{graphdist}.

\section{Bayesian Variable Selection for Cox Regression Model with Spatially Varying Coefficients}\label{sec:cox}

In this paper, we extend the spatially varying coefficient idea to the Cox regression model \citep{cox1972regression}. Let $t$ be the survival time of a subject in the site $s$ and $h(t(s)|\bm{x}(s))$ denote its hazard function for the subject with covariate vector $\bm{x}(s)$. Then we consider the Cox model for the subject in each site $s$,
\begin{equation}
\label{cox}
    h(t(s)|\bm{x}(s))=h_0(t(s))\exp(\bm{x}^T(s)\bm{\beta}(s)),
\end{equation}
where $h(\cdot)$ is the hazard function with the baseline hazard $h_0(\cdot)$, 
and $\bm{\beta}(s)=(\beta_1(s),\ldots,\beta_p(s))^T$ are the $p$-dimensional regression coefficients. For different sites, the regression model can be different since the regression coefficients are allowed to vary spatially.
Assuming no ties are present among the survival times for each site. Let $t_1(s)< t_2(s) <\ldots < t_{D(s)}(s)$ denote the ordered survival times from the site $s$ with $\bm{x}_{(j)}(s)$ being the covariate vector associated with
the individual who dies at time $t_j(s)$, and $\mathcal{R}(t_j(s))$ denote the set of all individuals at risk prior to $t_j$ at site $s$. 
Then the partial likelihood function for the site $s$ is written as
\begin{equation}
    L(\bm{\beta}(s)) = \prod_{j=1}^{D(s)} \frac{\exp(\bm{x}^T_{(j)}(s)\bm{\beta}(s))}
    {\sum_{l\in\mathcal{R}(t_j(s))}\exp(\bm{x}^T_l(s)\bm{\beta}(s))}.
    \label{plikely}
\end{equation}
So we can maximize each partial likelihood function per site using existing software, and the corresponding estimator is called the partial maximum likelihood estimator (PMLE).

Like the usual regression models, not all the predictors have significant effects on hazards or survival times, so it would be worthwhile to develop a variable selection method for the above model. Furthermore, for the covariates selected in the model, their effects may differ among locations. Therefore, finding out which variables are spatially varying can make \eqref{cox} more accurate and more flexible.

For the spatially varying coefficients model, we need to carry out variable selection in two different levels. For the first level, we need to select the variables over the whole space. That means we need to determine which variables are significant for all the locations. The second level is to determine which variables have spatially varying effects. For the significant variable selection in the first level, we extend the horseshoe prior by including a spatial correlation matrix for spatially varying coefficient models. Suppose there are $n$ sites, $s_1,\ldots,s_n$. Then the horseshoe version of \eqref{GL mixturess} can be extended to a multidimensional horseshoe prior as:
\begin{equation}
		\bm{\beta}_k \sim \text{N}_n(\bm{0}_n,\tau^2\lambda^2_k \bm{H}_k),\qquad
		\tau \sim \text{Ca}^+(0,1),\quad
		\lambda_k \sim \text{Ca}^+(0,1),
	 \label{horseshoe}
 \end{equation}
where $\bm{\beta}_k=(\beta_k(s_1),....,\beta_k(s_n))^T$ is the $k$-th regression coefficient vector across all sites, $k=1,\ldots,p$, and $\bm{H}_k$ is the corresponding spatial correlation matrix. $\tau^2$ controls the global variance for all the regression coefficients, and $\lambda_k^2$ is the local variance parameter for the $k$-th regression coefficient vector.

Note $\bm{H}_k,$ a $n$-by-$n$ matrix, models the spatial variation of  $\bm{\beta}_k$.
We define the $(l,m)$-th entry of $\bm{H}_k$ by  
\begin{equation}
    \bm{H}_k^{(l,m)} = \exp\{-\gamma_k\times d(s_l, s_m)\},
    \label{hmatrix}
\end{equation}
where $d(s_l,s_m)$ is the graph distance \eqref{graphdist} between site $l$ and site $m$. In this way, the correlation of $\beta_k(s_l)$ and $\beta_k(s_m)$ depends on the distance between these two sites, and for a fixed $\gamma_k$, the correlation gets smaller if two sites are more distant. The multidimensional horseshoe prior can deal with the multidimensional variable selection problems, especially when the regression coefficients are spatially dependent. In order to determine which coefficients are spatially varying over the space, we give the point mass mixture prior on $\gamma_k$ as follows:
\begin{equation}
\begin{split}
		\gamma_k|\pi_k \sim (1-\pi_k) \delta_0+\pi_k\text{Gamma}(a_0,b_0),\\
		\pi_k \sim \text{Beta}(0.5,0.5),
\end{split}
\label{gammaprior}
\end{equation}
where $\delta_0$ is the point with mass 1 at 0, and $a_0$ and $b_0$ are the shape and rate parameters of the gamma distribution with mean $a_0/b_0$. If $\gamma_k=0$, $\exp\{-\gamma_k\times d(s_l, s_m)\}=1$ for any $(s_l, s_m)$, and then $\beta_k(s_i)$'s are perfectly positive correlated across sites. In this case, $\bm{\beta}_k$ is regarded as spatially static since it doesn't depend on the distance $d(s_l,s_m)$. If $\gamma_k\neq0$, then $\exp\{-\gamma_k\times d(s_l, s_m)\}\neq1$, so $\beta_k(s_l)$ and $\beta_k(s_m)$ are correlated but can vary as the distance changes. Thus, $\bm{\beta}_k$ is considered to be spatially varying.

Combining \eqref{cox}, \eqref{horseshoe}, \eqref{hmatrix} and \eqref{gammaprior}, we can have the following hierarchical Bayesian variable selection model for subject $j$ in site $i$, $j=1,\ldots,n_i$ and $i=1,\ldots,n$.
\begin{equation}
	\begin{split}
		h(t_j(s_i)|\bm{x}_j(s_i))= h_0(t_j(s_i)) \exp\{\bm{x}^T_j(s_i)\bm{\beta}(s_i)\}\\
		\bm{\beta}_k|\tau, \lambda_k, \bm{H}_k \sim \text{N}_n(\bm{0}_n,\tau^2\lambda^2_k \bm{H}_k)\\
		\tau \sim \text{Ca}^+(0,1)\\
		\lambda_k \sim \text{Ca}^+(0,1)\\
		\bm{H}_k^{(l,m)}|\gamma_k = \exp\{-\gamma_k\times d(s_l, s_m)\}\\
		\gamma_k|\pi_k \sim (1-\pi_k) \delta_0+\pi_k\text{Gamma}(a_0,b_0)\\
		\pi_k \sim \text{Beta}(0.5,0.5)
	 \end{split}
	 \label{svc variable selection}
 \end{equation}

In the posterior inference, a regression coefficient is regarded as not significant if the posterior mean of the corresponding local shrinkage parameter $\lambda_k$ is less than 1 according to \cite{carvalho2009handling,carvalho2010horseshoe}. For spatially varying coefficient detection, if $\gamma_k = 0$ occurs with the posterior probability greater than $0.5$, the corresponding regression coefficient $\bm{\beta}_k$ is considered to be not spatially varying. Otherwise, it will be regarded as spatially varying.

\section{Bayesian Computation}\label{sec:est}
\subsection{Two-Stage Estimation Method}
In Bayesian framework, the lack of conjugacy makes it challenging to draw posterior samples for the regression coefficients in the Cox models. Considering the model complexity, we modify the
``two-stage approximation'' method proposed by
\cite{boehm2015spatial} to estimate the spatially varying coefficients. In the first stage, we obtain the PMLE for every site separately by maximizing the corresponding partial likelihood \eqref{plikely}. Then the original partial likelihood can be approximated by a normal likelihood using Taylor expansion and the Fisher information matrix:
\begin{equation} 	
  \hat{\bm{\beta}}(s_i)\sim \text{N}(\bm{\beta}(s_i),\hat{\bm{V}}(s_i)),
	\label{approximationlike}
\end{equation}
where $\hat{\bm{\beta}}(s_i)$ are PMLEs obtained from the data at location $s_i$, and $\hat{\bm{V}}(s_i)$ is their estimated covariance matrix which summarizes the variation of the data propagated to the regression coefficients.

In the second stage,  we set up a hierarchical structure on the $\bm{\beta}(s_i)$'s in \eqref{approximationlike} instead of the $\bm{\beta}(s_i)$'s in the original survival model. In this way, the $\hat{\bm{\beta}}(s_i)$'s are regarded as the data sampled from multivariate normal distributions, and the $\bm{\beta}(s_i)$'s are the mean vectors of these multivariate normal distributions where the covariance matrices are known and they are basically the estimated covariance matrices of the PMLEs. Then \eqref{svc variable selection} can be rewritten as:
\begin{equation}
	\begin{split}
		\hat{\bm{\beta}}(s_i)|\bm{\beta}(s_i) \sim \text{N}(\bm{\beta}(s_i),\hat{\bm{V}}(s_i))\\
		\bm{\beta}_k |\tau, \lambda_k, \bm{H}_k \sim \text{N}_n(\bm{0}_n,\tau^2\lambda^2_k \bm{H}_k)\\
		\tau \sim \text{Ca}^+(0,1)\\
		\lambda_k \sim \text{Ca}^+(0,1)\\
		\ldots
		\label{hierarchical}
	 \end{split}
 \end{equation}
 where $\bm{\beta}_k=(\beta_k(s_1),....,\beta_k(s_n))^T$ is the $k$-th regression coefficient vector as before.
 
In the two-stage estimation method, the partial likelihood functions are approximated by normal likelihoods, so we only need to maximize the partial likelihood once in the first stage. Then we can build an MCMC algorithm based on \eqref{hierarchical}
 and such an algorithm is much more computationally efficient than that based on the original hierarchical model \eqref{svc variable selection}.

Note the error propagated using the two stage estimation procedure is not the same as the original hierarchical model given a $\hat{\bm{\beta}}(s_i)$ is introduced. However, as the sample size in each site becomes large, we expect our two-stage procedure to work well, because the $\bm{\beta}(s_i)$ here can be thought of as the same as the $\bm{\beta}(s_i)$ in the Cox model. A reviewer suggested two different approaches for a finite sample size. One is to propagate uncertainty using multiple simulations of the results of stage 1 \citep{blangiardo2016two, liu2017incorporating}. The other is to propagate uncertainty considering the result of stage 1 as a random variable in stage 2 with an informative prior obtained from the fit of stage 1 \citep{warren2012spatial, lee2017rigorous}.

The corresponding computing method and MCMC algorithm are implemented via \textsf{R} package \pkg{nimble} \citep{nimble}. The \pkg{nimble} code is listed in the appendix.

\subsection{Choice of Hyperprior}
In practice, we find it is challenging to specify the hyperprior $\text{Gamma}(a_0, b_0)$. It may inflate the false discovery rate if the mean of $\text{Gamma}(a_0, b_0)$ is small while the variance is relatively large, and it will inflate false omission rate if the mean is large while the variance is small. Therefore, it is appropriate to choose a gamma distribution with a moderate mean value and relatively small variance. In this paper, we choose $a_0=25$ and $b_0=50$ and it seems work well in practice regarding the balance of sensitivity and specificity . Thus, \eqref{gammaprior} can be rewritten as
\begin{equation}
\begin{split}
		\gamma_k|\pi_k \sim (1-\pi_k) \delta_0+\pi_k\text{Gamma}(25, 50),\\
		\pi_k \sim \text{Beta}(0.5,0.5).
\end{split}
\label{gammaprior2}
\end{equation}

In order to show the impacts of the hyperprior on the results, we also explore other hyperpriors in the simulation studies. Those hyperpriors include $\text{Gamma}(2.5,5)$, $\text{Gamma}(250,500)$,
$\text{Gamma}(16,40)$ and $\text{Gamma}(36,60)$.
Compared to $\text{Gamma}(25,50)$, $\text{Gamma}(2.5,5)$ and $\text{Gamma}(250,500)$ have the same mean but relatively large and small variances respectively, and $\text{Gamma}(16,40)$ and $\text{Gamma}(36,60)$ have the same variance but relatively small and large mean values respectively.

\section{Simulation Study}\label{sec:simulation}
We conducted three simulation studies to illustrate the two-stage variable selection method. In all simulated data sets, we consider $p=20$, the number of sites $n=64$ and generate each predictor from a standard normal distribution. In the first simulation study, we set the first ten components of $\bm{\beta}$'s to be zero vectors, that is, $\bm{\beta}_k=\bm{0}_n, k=1,\ldots,10$, so the first ten predictors are not expected to be selected in the model. The next five components  $\bm{\beta}_k, k=11,\ldots,15$, are set as spatially stationary, and they are equal to $(k-10)\cdot\bm{1}_n$, respectively. The last five components are spatially varying coefficients and the corresponding vectors $\bm{\beta}_k$'s are simulated from a multivariate normal distribution. For this multivariate normal distribution, each marginal is $N(3,1)$ and the correlation structure is based on the geographical information of Louisiana counties. For example, the $(l,m)$-th entry of the correlation matrix is $\exp\{-10\times d(s_l,s_m)\}$, where $d(s_l,s_m)$ is the graph distance between the $l$-th county and the $m$-th county. The sample size of each county is set to be 100. 
For each site, survival times are generated based on a Cox model with a constant baseline hazard function to be 0.5 and censored at a fixed time (155 in our example). The average censoring rate is around 35\%.

The above simulation process was repeated for 100 times, and in each simulation replicate, we apply the two stage estimation method. The MCMC chain in the second stage ran for 1,000,000 iterations with the first 900,000 as burn-in. The thinning factor was 20  to improve the independence of the posterior samples. The final posterior Monte Carlo sample size was 5,000 and the variable selection capability of our model is measured using true positive rate (TPR), true negative rate (TNR), positive predictive value (PPV) and negative predictive value (NPV). The formulas are listed as follows:
\begin{align*}
\mathrm {TPR} &={\frac {\mathrm {TP} }{\mathrm {TP} +\mathrm {FN} }}, &
\mathrm {TNR} &={\frac {\mathrm {TN} }{\mathrm {TN} +\mathrm {FP} }},\\
\mathrm {PPV} &={\frac {\mathrm {TP} }{\mathrm {TP} +\mathrm {FP} }}, &
\mathrm {NPV} &={\frac {\mathrm {TN} }{\mathrm {TN} +\mathrm {FN} }}.
\end{align*}
Both variable selection procedures are evaluated by the above four criteria. For significant variable selection, all of the 20 predictors are considered. In this case,
TP is the number of the significant predictors selected in the model while TN is the number of predictors with no effects excluded from the model.
FP and FN are the number of predictors with no effects but selected into the model and the number of significant predictors that are not selected into the model, respectively. Here, we must have TP+FN=TN+FP=$10$. 
In the detection of spatially varying predictors, only the last 10 predictors are under consideration since it is meaningless to find a non-significant regression coefficient to be spatially varying.
TP and TN are the number of spatially varying predictors that are being regarded as spatially varying correctly and the number of  spatially stationary predictors that are not being regarded as spatially varying, respectively;
while FP and FN are the number of spatially stationary predictors that are regarded as spatially varying and the number of spatially varying predictors that are not being detected correctly. In this evaluation, TP+FN=TN+FP=$5$.

Table \ref{table1} displays the simulation results discussed so far. The middle column displays the average operating characteristics for detecting a significant covariate, while the right column presents the average operating characteristics for detecting spatially varying coefficients. Corresponding standard deviations are listed in parentheses. For different hyperpriors, Gamma(25,50) can yield a good balance between TPR and TNR. Both TPR and TNR decrease if a hyperprior with a larger variance is selected, while TPR decreases but TNR increases if a smaller variance being selected. Regarding different mean values in the hyperprior, a smaller mean value will lead to a lower TNR, and a larger mean value will lead to a lower TPR. We can see our proposed method can do well in significant variable selection. For spatially varying coefficient detection, TPR and TNR are also good and both are over 80\% when Gamma(25, 50) is selected, but PPV and NPV are not able to be calculated sometimes since all significant variables are regarded as spatially varying or spatially static in some simulation replicates. These findings are supported by Table \ref{table2} as well, which reports the frequencies in 100 replications of a predictor being selected and a selected predictor being spatially varying for the Gamma(25, 50) hyperprior. $\bm{\beta}_{1}$ to $\bm{\beta}_{20}$ can be selected and detected correctly at both levels for most of the times except $\bm{\beta}_{16}$ that fails to be detected as spatially varying in 44 simulation replicates.

\begin{table}[h!]
\caption{Evaluation results of the first simulation study for the regular situation with 100 replications (standard deviation for each measure is given in the  parentheses)}
\label{table1}
\center
\begin{tabular}{p{2.8cm}|p{3cm}|p{3.5cm}|p{4cm} }
\hline
 Operating characteristics & Hyperprior & For detecting significant  predictor	& For detecting spatially varying coefficient \\
 \hline
\multirow{3}{*}{TPR \%} & Gamma(25,50) & 99.0(4.14) & 82.2(21.44)\\
& Gamma(2.5,5) & 99.3(2.93) & 72.8(23.36)\\ 
& Gamma(250,500) & 98.8(3.83) & 79.0(19.15)\\ 
& Gamma(16,40) & 99.6(2.43) & 83.0(19.15)\\
& Gamma(36,60) & 99.1(3.21) & 69.4(23.17)\\\hline
\multirow{3}{*}{TNR \%} & Gamma(25,50) & 99.9(1.00) & 93.2(21.88)\\
& Gamma(2.5,5) & 100(0) & 90.2(26.59)\\ 
& Gamma(250,500) & 100(0) & 100(0)\\ 
& Gamma(16,40) & 99.2(4.86) & 86.8(30.91) \\
& Gamma(36,60) & 99.9(1.00) & 98.2(12.74)\\\hline
\multirow{3}{*}{PPV \%} & Gamma(25,50) & 99.9(0.91) & NaN(NA)\\
& Gamma(2.5,5) & 100(0) & NaN(NA)\\ 
& Gamma(250,500) & 100(0) & 100(0)\\ 
& Gamma(16,40) & 99.4(3.67) & NaN(NA)\\
& Gamma(36,60) & 99.9(0.91) & NaN(NA)\\\hline
\multirow{3}{*}{NPV \%} & Gamma(25,50) & 99.1(3.41) & NaN(NA)\\
& Gamma(2.5,5) & 99.4(2.57) & NaN(NA)\\ 
& Gamma(250,500) & 98.9(3.34) & 84.6(12.52)\\
& Gamma(16,40) & 99.7(2.09) & NaN(NA)\\
& Gamma(36,60) & 99.2(2.83) & NaN(NA)\\\hline
\end{tabular}
\end{table}

\begin{table}[h!]
\caption{Frequency results of the first simulation study for the regular situation with 100 replications for each $\bm{\beta}$ with hyperprior Gamma(25, 50)}
\label{table2}
\center
\begin{tabular}{p{0.5cm}|p{1.8cm}|p{3cm}|p{0.5cm}|p{1.8cm}|p{3cm}}
\hline
$\bm{\beta}_k$ & $\bm{\beta}_k$ being selected & $\bm{\beta}_k$ detected as spatially varying & $\bm{\beta}_k$ & $\bm{\beta}_k$ being selected & $\bm{\beta}_k$ detected as spatially varying \\
\hline
$\bm{\beta}_1$ & 0 & (not applicable) & $\bm{\beta}_{11}$ & 93 & 8\\
$\bm{\beta}_2$ & 0 & (not applicable) & $\bm{\beta}_{12}$ & 98 & 5\\
$\bm{\beta}_3$ & 0 & (not applicable) & $\bm{\beta}_{13}$ & 99 & 7\\
$\bm{\beta}_4$ & 0 & (not applicable) & $\bm{\beta}_{14}$ & 100 & 7\\
$\bm{\beta}_5$ & 0 & (not applicable) & $\bm{\beta}_{15}$ & 100 & 7\\
$\bm{\beta}_6$ & 0 & (not applicable) & $\bm{\beta}_{16}$ & 100 & 56\\
$\bm{\beta}_7$ & 1 & 1 & $\bm{\beta}_{17}$ & 100 & 91\\
$\bm{\beta}_8$ & 0 & (not applicable) & $\bm{\beta}_{18}$ & 100 & 96\\
$\bm{\beta}_9$ & 0 & (not applicable) & $\bm{\beta}_{19}$ & 100 & 96\\
$\bm{\beta}_{10}$ & 0 & (not applicable) & $\bm{\beta}_{20}$ & 100 & 72\\
\hline
\end{tabular}
\end{table}

We use the average mean squared error (MSE) over all  sites to evaluate the accuracy of  the estimator of a regression coefficient. The MSE is first calculated for each regression coefficient by site, and then averaged across all 64 sites. The results with hyperprior Gamma(25, 50) are presented in Table \ref{table_accuracy}. For the non-significant variables, the estimates of the regression coefficients are very accurate, while for the significant ones, the average MSE gets larger as the magnitude of the true value increases.

\begin{table}[h!]
\caption{Accuracy evaluation for the estimators of the regression coefficients with hyperprior Gamma(25, 50)}
\label{table_accuracy}
\center
\begin{tabular}{p{1cm}|p{3cm}|p{1cm}|p{3cm}}
\hline
$\bm{\beta}$ & Average MSE & $\bm{\beta}$ & Average MSE \\
\hline
$\bm{\beta}_1$ & $<0.001$ & $\bm{\beta}_{11}$ & 0.222\\
$\bm{\beta}_2$ & $<0.001$ & $\bm{\beta}_{12}$ & 0.865\\
$\bm{\beta}_3$ & $<0.001$ & $\bm{\beta}_{13}$ & 1.936\\
$\bm{\beta}_4$ & $<0.001$ & $\bm{\beta}_{14}$ & 3.429\\
$\bm{\beta}_5$ & $<0.001$ & $\bm{\beta}_{15}$ & 5.343\\
$\bm{\beta}_6$ & $<0.001$ & $\bm{\beta}_{16}$ & 2.209\\
$\bm{\beta}_7$ & $<0.001$ & $\bm{\beta}_{17}$ & 2.163\\
$\bm{\beta}_8$ & $<0.001$ & $\bm{\beta}_{18}$ & 1.993\\
$\bm{\beta}_9$ & $<0.001$ & $\bm{\beta}_{19}$ & 2.188\\
$\bm{\beta}_{10}$ & $<0.001$ & $\bm{\beta}_{20}$ & 2.223\\
\hline
\end{tabular}
\end{table}

To further illustrate our proposed model, we modify the sparsity patterns and spatial patterns in simulation study 2 by setting only the 19th and 20th regression coefficients as significant and only the 20th regression coefficient as spatially varying. Here $\bm{\beta}_{19}=3\cdot\bm{1}_n$, and $\bm{\beta}_{20}\sim \text{N}_{n}(3\cdot \bm{1}_{n}, \exp\{-10\times \bm{D}\})$ where $\bm{D}$ is the distance matrix with $(l,m)$-th entry as $d(s_l,s_m)$. All other $\bm{\beta}_k = \bm{0}_n$. In this case, it is not worthwhile to look at the operating characteristics since there are only two significant predictors and only one of them has spatially varying effects. The frequency results are presented in Table \ref{table3}. Under the very sparse pattern, the significant variables can always be selected, but some non-significant variables sometimes are also selected into the model. Once they are selected, they are almost always detected to be spatially varying. In our model, there are two types of variation; variation from individuals within each site and variation across sites. We think in this scenario only two significant predictors are not sufficient to explain all the variation so some false positives are expected.

\begin{table}[h!]
\caption{Frequency results of simulation study 2 for the extreme sparse situation with 100 replications for each $\bm{\beta}$ with hyperprior Gamma(25, 50)}
\label{table3}
\center
\begin{tabular}{p{0.5cm}|p{1.8cm}|p{3cm}|p{0.5cm}|p{1.8cm}|p{3cm}}
\hline
$\bm{\beta}_k$ & $\bm{\beta}_k$ being selected & $\bm{\beta}_k$ detected as spatially varying & $\bm{\beta}_k$ & $\bm{\beta}_k$ being selected & $\bm{\beta}_k$ detected as spatially varying \\
\hline
$\bm{\beta}_1$ & 43 & 43 & $\bm{\beta}_{11}$ & 45 & 45\\
$\bm{\beta}_2$ & 40 & 40 & $\bm{\beta}_{12}$ & 40 & 40\\
$\bm{\beta}_3$ & 37 & 37 & $\bm{\beta}_{13}$ & 41 & 41\\
$\bm{\beta}_4$ & 53 & 53 & $\bm{\beta}_{14}$ & 43 & 43\\
$\bm{\beta}_5$ & 38 & 38 & $\bm{\beta}_{15}$ & 50 & 49\\
$\bm{\beta}_6$ & 44 & 44 & $\bm{\beta}_{16}$ & 44 & 43\\
$\bm{\beta}_7$ & 34 & 34 & $\bm{\beta}_{17}$ & 34 & 34\\
$\bm{\beta}_8$ & 41 & 40 & $\bm{\beta}_{18}$ & 47 & 47\\
$\bm{\beta}_9$ & 47 & 47 & $\bm{\beta}_{19}$ & 100 & 1\\
$\bm{\beta}_{10}$ & 31 & 31 & $\bm{\beta}_{20}$ & 100 & 100\\
\hline
\end{tabular}
\end{table}

In simulation study 3, we keep the same sparsity patterns and spatial patterns as in simulation study 1 but adjust the spatial signals. For spatially varying coefficients, the spatial correlation matrix is set as $\exp\{-\bm{D}\}$ instead of $\exp\{-10\times \bm{D}\}$, which means the spatial effects are weaker than those in the simulation study 1. Corresponding results are gathered in Table \ref{table4} and \ref{table5}. In this simulation study, hyperprior Gamma(25, 50) still works well and reaches a balance between TPR and TNR, and when hyperprior Gamma(25, 50) is selected, our model works well in significant variable selection but the performance is not that robust in spatially varying coefficient detection. TPR, TNR and PPV values are good but the standard deviation of TNR is large, and NPV can not be calculated since the significant regression coefficients are all regarded as spatially varying in some simulation replicates.

\begin{table}[h!]
\caption{Evaluation results of the simulation study 3 with weak spatial signal with 100 replications (standard deviations are in parentheses)}
\label{table4}
\center
\begin{tabular}{p{2.8cm}|p{3cm}|p{3.5cm}|p{4cm} }
\hline
 Operating characteristics & Hyperprior & For detecting  significant predictor	& For detecting spatially varying predictor \\
 \hline
\multirow{3}{*}{TPR \%} & Gamma(25,50) & 98.9(4.69) & 88.0(13.63)\\
& Gamma(2.5,5) & 80.7(7.56) & 0(0)\\ 
& Gamma(250,500) & 97.9(6.40) & 81.6(15.22)\\ 
& Gamma(16,40) & 98.4(6.77) & 89.8(12.55)\\
& Gamma(36,60) & 99.1(4.04) & 81.0(15.41)\\\hline
\multirow{3}{*}{TNR \%} & Gamma(25,50) & 99.7(1.71) & 84.8(32.05)\\
& Gamma(2.5,5) & 100(0) & 0(0)\\ 
& Gamma(250,500) & 100(0) & 99.8(2.00)\\ 
& Gamma(16,40) & 99.4(3.12) & 74.8(39.96) \\
& Gamma(36,60) & 99.7(1.71) & 96.4(16.91)\\\hline
\multirow{3}{*}{PPV \%} & Gamma(25,50) & 99.7(1.56) & 91.1(17.42)\\
& Gamma(2.5,5) & 100(0) & NaN(NA)\\ 
& Gamma(250,500) & 100(0) & 99.8(2.00)\\ 
& Gamma(16,40) & 99.5(2.65) & 86.1(20.94)\\
& Gamma(36,60) & 99.7(1.56) & 97.9(9.33)\\\hline
\multirow{3}{*}{NPV \%} & Gamma(25,50) & 99.1(3.64) & NaN(NA)\\
& Gamma(2.5,5) & 84.2(5.31) & NaN(NA)\\ 
& Gamma(250,500) & 98.3(5.06) & 85.8(10.94)\\
& Gamma(16,40) & 98.7(4.70) & NaN(NA)\\
& Gamma(36,60) & 99.2(3.31) & NaN(NA)\\\hline
\end{tabular}
\end{table}

\begin{table}[h!]
\caption{Frequency results of the simulation study 3 with weak spatial signal with 100 replications for each $\bm{\beta}$}
\label{table5}
\center
\begin{tabular}{p{0.5cm}|p{1.8cm}|p{3cm}|p{0.5cm}|p{1.8cm}|p{3cm}}
\hline
$\bm{\beta}_k$ & $\bm{\beta}_k$ being selected & $\bm{\beta}_k$ detected as spatially varying & $\bm{\beta}_k$ & $\bm{\beta}_k$ being selected & $\bm{\beta}_k$ detected as spatially varying \\
\hline
$\bm{\beta}_1$ & 1 & 1 & $\bm{\beta}_{11}$ & 92 & 12\\
$\bm{\beta}_2$ & 0 & (not applicable) & $\bm{\beta}_{12}$ & 99 & 15\\
$\bm{\beta}_3$ & 0 & (not applicable) & $\bm{\beta}_{13}$ & 99 & 17\\
$\bm{\beta}_4$ & 0 & (not applicable) & $\bm{\beta}_{14}$ & 99 & 15\\
$\bm{\beta}_5$ & 0 & (not applicable) & $\bm{\beta}_{15}$ & 100 & 16\\
$\bm{\beta}_6$ & 1 & 1 & $\bm{\beta}_{16}$ & 100 & 51\\
$\bm{\beta}_7$ & 1 & 1 & $\bm{\beta}_{17}$ & 100 & 100\\
$\bm{\beta}_8$ & 0 & (not applicable) & $\bm{\beta}_{18}$ & 100 & 89\\
$\bm{\beta}_9$ & 0 & (not applicable) & $\bm{\beta}_{19}$ & 100 & 100\\
$\bm{\beta}_{10}$ & 0 & (not applicable) & $\bm{\beta}_{20}$ & 100 & 100\\
\hline
\end{tabular}
\end{table}

\section{Data Analysis}\label{sec:realdata}
\begin{figure}[h!]
  \centering
    \includegraphics[width=\linewidth]{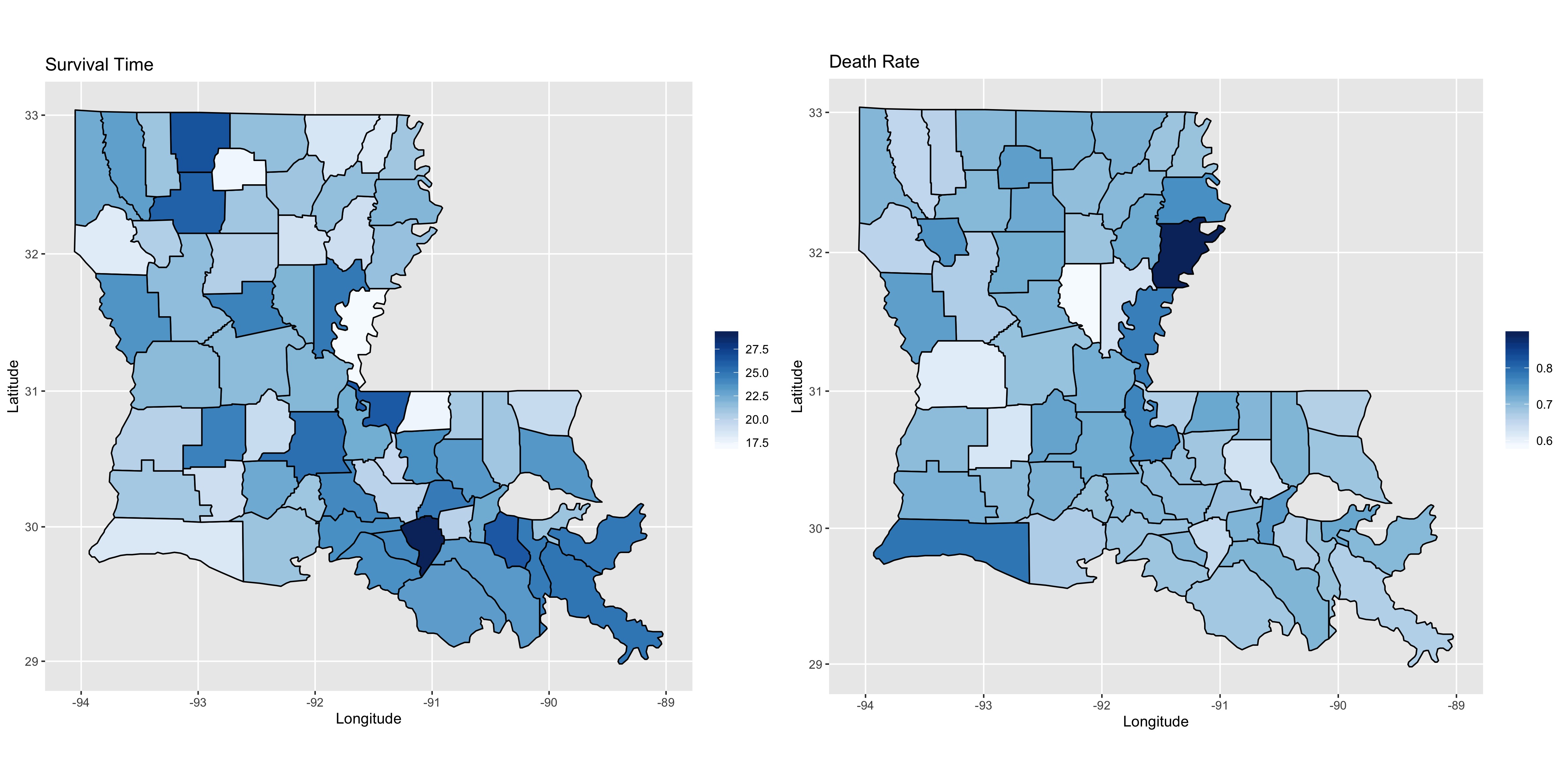}
  \caption{Maps of spatial patterns of mean survival times and death rates for all 64 counties.}
  \label{summary_map}
\end{figure}

In this section, we applied our proposed method to analyzing a respiratory cancer data set in Louisiana state, which was downloaded from the Surveillance, Epidemiology, and End Results (SEER) Program (\url{https://seer.cancer.gov/}). Modifying the criteria in \cite{zhang2016bayesian}, we excluded: 1) subjects for whom the respiratory cancer was not the primary cancer; 2) subjects with unknown marital status; 3) subjects with unknown race and race other than Black and White; 4) subjects with unknown sex; 5) subjects with unknown age at diagnosis; 6) subjects with unknown cancer stage; 7) subjects with unknown cancer grade; 8) subjects with unknown surgery status; 9) subjects with unknown radiation status; 10) subjects with unknown survival times; and 11) subjects who died not because of respiratory cancer. After cleaning, there were 16213 observations left. Figure \ref{summary_map} displays the mean survival times and death rates for each county. It can be observed that there are large variations in mean survival times and death rates across counties.

\begin{figure}[h!]
  \centering
    \includegraphics[width=\linewidth]{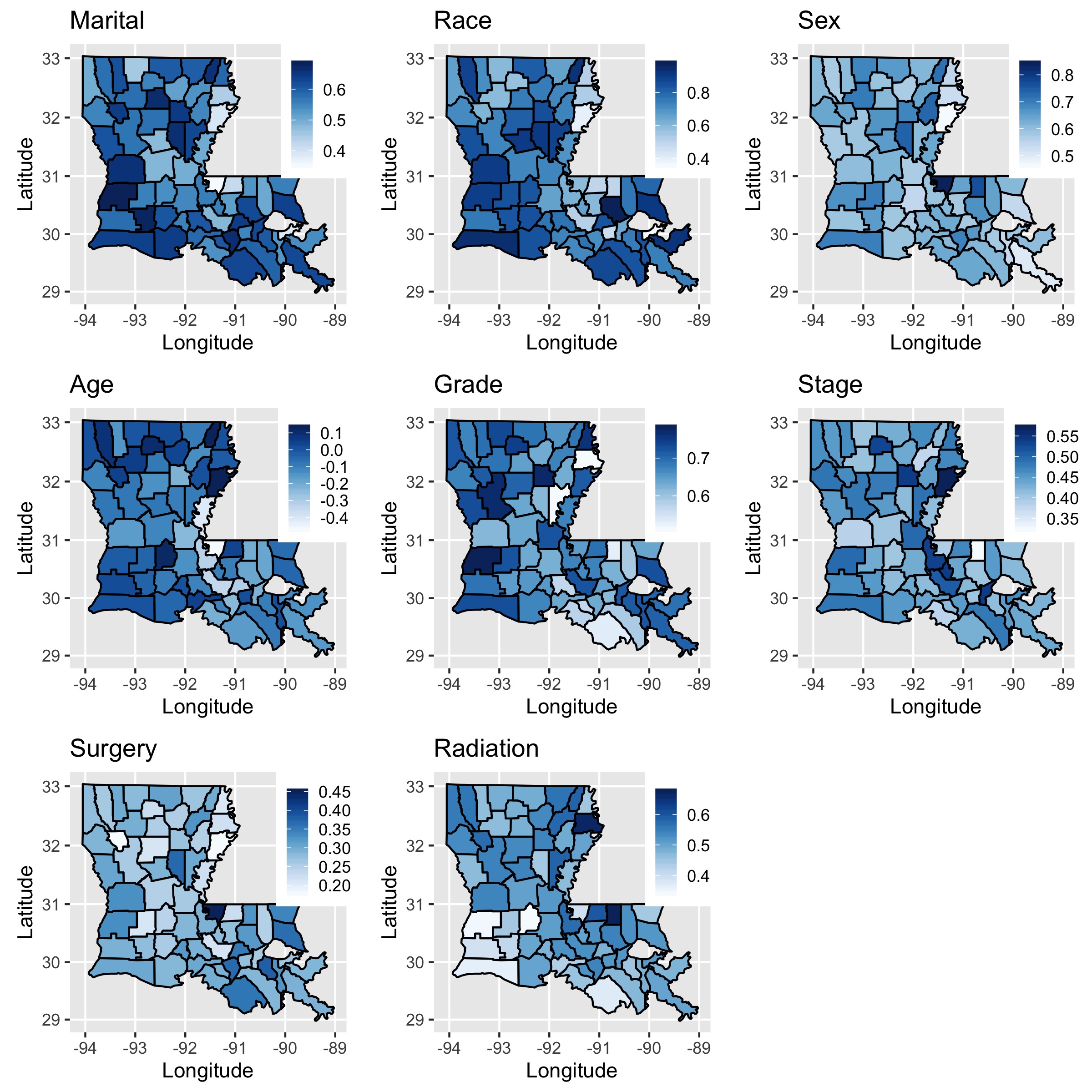}
  \caption{Maps of spatial patterns of each predictor. Color depth represents the magnitude of the proportion of married, white, male, distant, grade III or IV, surgery (yes) and radiation (yes) in each predictor except age. Age is a standardized continuous variable.}
  \label{predictor_map}
\end{figure}

We have selected 8 variables as potential covariates from the data set. Some demographic information are available including age at diagnosis, sex (male or female) and marital status (married or not). For race, we only include the black and white since the sample sizes for other races are small. Covariate information on surgery status (yes or no) and radiation status (yes or no) are also available and thus included. As in \cite{wu2015empirical}, we also include cancer stage and tumor grade as predictors and dichotomize them as distant or not, III or IV versus other grades, respectively. The spatial patterns of each predictor are presented in Figure \ref{predictor_map}. We can observe all the predictors have some spatial variations across different counties.

As introduced in Section 4, we first employed the Cox regression model for the data for each location. In Cameron Parish, there were only 2 white patients, resulting in inflated estimate of regression coefficient for race. Therefore, we excluded Cameron Parish from our data and 16180 observations were left. We applied our proposed model to the cleaned data set and the corresponding MCMC chain ran for 2,000,000 iterations. The burn-in number was 1,800,000 and the thinning factor was 20, so a total of 10,000 samples were used for final inference.

\begin{figure}[h!]
  \centering
    \includegraphics[width=\linewidth]{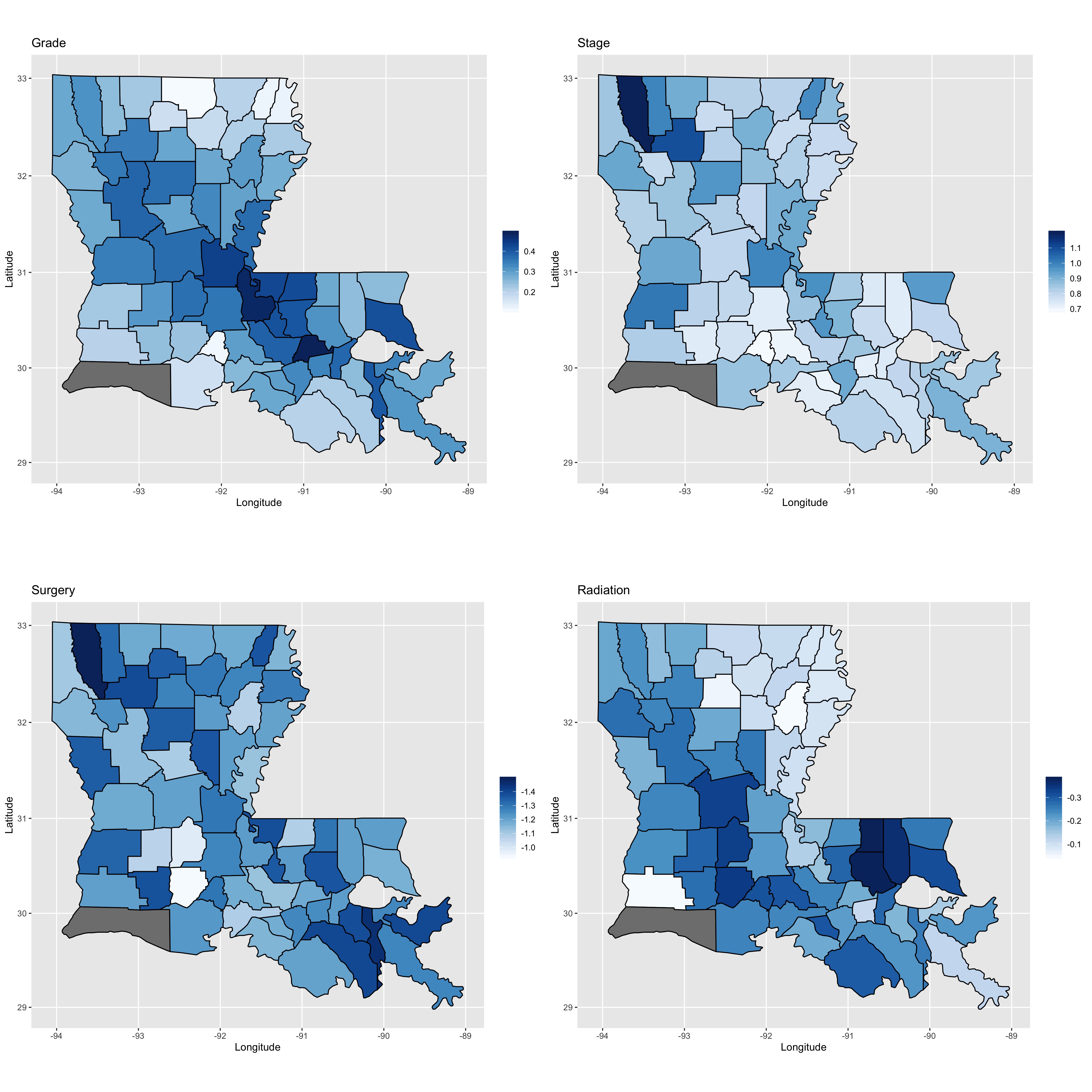}
  \caption{Maps of spatial patterns of the estimated values of regression coefficients for grade, stage, surgery status and radiation status. Cameron Parish was excluded for inference and was plotted in grey.}
  \label{beta_map}
\end{figure}

We find there are 4 variables selected into the model, including grade, stage, surgery status and radiation status, which means these four predictors have significant effects on survival times. Among these four significant predictors, grade and radiation status have spatially varying effects. Figure \ref{beta_map} shows the estimated values of regression coefficients of the four significant predictors. For the two spatially varying predictors, grade and radiation, we can find there are more significant spatial variation on the regression effects. Grade has larger regression effects in the central region than the surrounding areas. The low effects of radiation status mainly concentrate on a ribbon area, which is the Mississippi valley. The other two predictors, stage and surgery, were selected to be spatially stationary predictors. The results are also reasonable since the effects of stage and surgery appear to have less spatial variation, and more importantly, their variation seems spatially independent. The magnitude of the regression coefficients are randomly distributed over the whole space, instead of depending on the distance between counties.

\section{Discussion}\label{sec:discussion}
In this paper, we proposed a Bayesian variable selection method for the Cox regression model with spatially varying coefficients. We have constructed a hierarchical model for variable selection, where horseshoe priors were considered on each local smoothing and the global smoothing parameters to control sparsity. Additionally, on determining whether a coefficient is spatially varying or spatially stationary, we consider a mixture prior of a point mass and a continuous density for each entry of the spatial correlation matrix. On the computation, we have proposed a more efficient algorithm to approximate the solution. On the first stage, we derive the maximum likelihood estimates for each regression coefficient per site independently using existing software, for example, \textsf{R} package \pkg{survival}. Then we develop an MCMC algorithm for the hierarchical variable selection model based on the PMLE of the first stage. The MCMC algorithm was implemented in \pkg{nimble} which  compiles simple \textsf{R}-like code into \textsf{C++} for speed. The computer code is given in the Appendix. 

We used three simulated data sets to illustrate our method and found that it could handle the sparsity very well when the sparsity was not that high and could also detect the spatially varying coefficients correctly in most of the situations. We have also applied our method to a real data set for respiratory cancer in Louisiana from the SEER program. Starting with eight predictors in the data set, we have selected four predictors including grade, stage, surgery status and radiation status, to be significant in the Cox regression model and two of them (grade and radiation status) were selected to be spatially varying.

The future work may involve in the improvement of the robustness of our method. As reflected from the simulation study, the performance of detecting spatially varying predictors varied a lot for different data sets. For some simulated data sets, the proposed model worked well while for some other data sets, there were some false negatives. In addition, the choice of the ``slab'' part of the hyperprior on $\gamma_k$ can also be studied in the future since it is intricately related to the performance of the detection for spatially varying coefficients. Furthermore, the current model can only do significant variable selection overall for the whole space and then determine whether the selected variables are spatially varying. It is worthwhile to consider a site-specific selection method which allows covariates to have non-zero betas in some locations and betas=0 in other locations for the same covariates.

\section*{Appendix}

In the appendix, the \pkg{nimble} code is listed to demonstrate the MCMC algorithm. With \pkg{nimble}, we can write our own code in \textsf{R} but in \textsf{BUGS} syntax and then \pkg{nimble} can compile our code into \textsf{C++}. Before building the MCMC with \pkg{nimble}, we first use \textsf{R} package \pkg{survival} to obtain the PMLEs and corresponding variance matrices.

\begin{lstlisting}
> dat$"survobj" <- with(dat, Surv(time, status == 1))
> surv.x <- paste("X", 1:20, sep = "")
> formula <- as.formula(paste("survobj ~ ", 
+                              paste(surv.x, 
+                                    collapse = "+")))
> betahat <- matrix(0, nrow = 64, ncol = 20)
> Vhat <- array(0, c(64, 20, 20))
> for(i in 1:64){
+   dat.i <- dat[dat$site.ind==i, ]
+   mod.i <- coxph(formula = formula, data = dat.i)
+   betahat[i, ] <- mod.i$coefficients
+   Vhat[i, , ] <- mod.i$var
+ }
\end{lstlisting}

Usually, a \pkg{nimble} model contains four parts: the model code, the constants, the data and the initial values. The first part is written with function \texttt{nimbleCode} and our model code is listed in the following:
\begin{lstlisting}
> spvs <- nimbleCode({
+   for (i in 1:N){
+     hatbeta[i,1:p] ~ dmnorm(beta[i,1:p], 
+                             cov = hatV[i,1:p,1:p])
+   }
+   for (i in 1:p){
+     beta[1:N,i] ~ dmnorm(mu_beta[1:N], 
+                          cov = beta_cov[i,1:N,1:N])
+     beta_cov[i,1:N,1:N] <- tau^2 * lambda[i]^2 *
+       exp(-gamma[i] * dist[1:N,1:N])
+     gamma[i] <- (1 - c[i]) * 0 + c[i] * gamma0[i]
+     gamma0[i] ~ dgamma(25, 50)
+     c[i] ~ dbern(pi[i])
+     pi[i] ~ dbeta(0.5,0.5)
\end{lstlisting}
The first \texttt{for} loop is the approximate multivariate normal distribution built on the PMLEs \texttt{hatbeta} and its estimated covariance matrix \texttt{hatV}. In the second \texttt{for} loop, the hierarchical structure on \texttt{beta} is defined as in \eqref{svc variable selection} and \eqref{gammaprior}. The half-Cauchy priors on global shrinkage parameter \texttt{tau} and local shrinkage parameter \texttt{lambda[i]} are given in the code below.
\begin{lstlisting}
+     # half-Cauchy prior for lambda_i
+     lambda[i] <- abs(alam[i] / blam[i])
+     alam[i] ~ dnorm(0,1)
+     blam[i] ~ dnorm(0,1)
+   }
+   # half-Cauchy prior for tau
+   tau <- abs(atau/btau)
+   atau ~ dnorm(0,1)
+   btau ~ dnorm(0,1)
+ })
\end{lstlisting}
After defining the model, we can specify the data, the constants and the initial values as follows.
\begin{lstlisting}
> data <- list(hatbeta = betahat, hatV = Vhat, 
+              dist = grap.dist)
> constants <- list(N = n, p = p, mu_beta = rep(0, n))
> inits <- list(c = rep(1,p),
+               gamma0 = rep(1,p),
+               atau = 0.1,
+               btau = 0.1,
+               alam = rep(0.1,p),
+               blam = rep(0.1,p))
\end{lstlisting}
Next we can combine the four parts together using function \texttt{nimbleModel} and build MCMC algorithm by specifying some configurations.
\begin{lstlisting}
> svvsModel <- nimbleModel(spvs,
+                          data = data,
+                          constants = constants,
+                          inits = inits,
+                          check = FALSE)
> survConf <- configureMCMC(model = svvsModel,
+                           monitors = c("beta", "lambda", 
+                                        "c"))
> survMCMC <- buildMCMC(conf = survConf)
\end{lstlisting}
In the above code, \texttt{monitors} refers to the parameters we mainly focus on and the corresponding posterior samples will be output.
The \pkg{nimble} model and corresponding MCMC algorithm can be compiled into \textsf{C++}.
\begin{lstlisting}
> survModel.C <- compileNimble(svvsModel)
> survMCMC.C <- compileNimble(survMCMC, 
+                             project = survModel.C)
\end{lstlisting}
Finally, we can run our MCMC algorithm and get posterior samples. In the following code, two MCMC chains are derived, the total number of iterations is 50000 with the first 40000 as burn-in and the thinning rate is 10.
\begin{lstlisting}
> mcmc.out <- runMCMC(mcmc = survMCMC.C,
+                     niter = 50000, nchains = 2, 
+                     nburnin = 40000,
+                     thin = 10, summary = TRUE)
\end{lstlisting}

\bibliographystyle{apalike}
\bibliography{main}   
\end{document}